\documentclass[11pt,a4paper]{article}
\usepackage{ifpdf}
\usepackage{ae}
\usepackage[T1]{fontenc}
\usepackage[utf8]{inputenc}
\usepackage{mathrsfs}
\usepackage{amsmath}
\usepackage{amssymb}
\pdfoutput=1
\usepackage{hyperref}
 \usepackage{
a4wide,graphicx,bm,times,psfrag,wrapfig,sidecap} \usepackage{cite}
\makeatletter \let\old@startsection=\@startsection
\renewcommand{\@startsection}[6]
{\old@startsection{#1}{#2}{#3}{#4}{#5}{#6\mathversion{bold}}}

\newcommand\re[1]{({\ref{#1}})} \def\be{\begin{eqnarray} }
\def\ee{\end{eqnarray}}

  \def\CZ{{ \mathcal{
Z} }}  
  
\def\bee{\be\begin{aligned}} \def\eee{\end{aligned} \ee }
 \def\la{\label}

\def\({\left(} \def\){\right)} \def\<{\langle\,} \def\>{\, \rangle}
   \def\CZ{{
\mathcal{ Z} }}  
 
 \def\M{ {M}} 
   \def\CC{{
\mathcal{ C} }}   
   \def\a{\alpha}

\def\be{\begin{eqnarray}} \def\ee{\end{eqnarray}} 
\def\la{\label}
\def\({\left(} \def\){\right)} \def\<{\left\langle\,} \def\>{\,
\right\rangle} \def\[{\left[} \def\]{\right]} 
 \def\tr{{\rm tr} }

\def\hf{ {\textstyle{1\over 2}} }   \def\CC{{ \mathcal{ C} }}
    
    \def\CZ{{ \mathcal{ Z} }}
 \def\X{\mathrm{X}}  
   \def\d{\delta} 
   \def\a{\alpha} 
  \def\d{\delta}  
   \def\s{\sigma} \def\t{\tau} 
   
          \def\hf{ \frac{1}{2}}

   \def\Y{{\mathrm{Y}} }
  \def\X{{ \mathrm{X}}}

\def\Vand{\Delta}

 \newcommand{\K}{  {N}}  
\newcommand{\Mbar}{\overline{\mathcal{M}}}
\newcommand{\HN}{\mathcal{H}_N}
 \newcommand{\FC}{ C}  
 \newcommand{\dd}{\mathrm{d}}

 \usepackage{amsmath, amssymb,amsfonts,times,graphics,psfrag}
 \usepackage{graphicx} \usepackage{amssymb}

\begin{document}
\vspace*{-.4in} \thispagestyle{empty} \vspace{.1in} {\Large
\begin{center}
  {\bf A Matrix Model for Higher-Genus Fuss--Catalan Numbers }
\end{center}}

\begin{center}
\vspace{8mm}

\begin{center}
Anatol Kirillov$^{\ast}$ and Ivan Kostov$^{\ast}$$^{\star}$ 
\\[7mm]

	 {\it $^{\ast}$ Beijing Institute of Mathematical Sciences and
	 Applications (BIMSA)\\
  Huairou 101408, Beijing, China \\[5mm]

{\it $^{\star}$ Universit\'e Paris-Saclay, CNRS, CEA, Institut de
Physique th\'eorique \\
	 91191 Gif-sur-Yvette, France } \\[5mm]}

\end{center}

\vskip9mm

\vskip18mm

\end{center}

 \vskip 1cm 

{The genus-$g$ Fuss--Catalan (FC) number counts the number
of ways to obtain a genus-$g$ surface by identifying the edges of a
$pn$-gon via $p$-valent hyperedges.  For $p=2$ these are the
genus-$g$ Catalan numbers which are generated as the trace
correlations in the Gaussian matrix model (GUE).  Here we construct a
simple two-matrix model which generates the higher-genus
Fuss-Catalan numbers for any $p$ as the coefficients of its
$1/N$-expansion.  We obtain exact sum rules and an explicit formula
for the higher-genus Fuss-Catalan numbers which generalises the
Harer-Zagier formula to $p>2$.  We discuss the relation of the
higher-genus FC numbers to the intersection numbers and the Euler
characteristic of the moduli space of spin-$p$ curves. }

 \vskip 1cm 

\newpage 
 
 \section{Introduction}

\subsection{ The classical Fuss-Catalan numbers}

The numbers now called \emph{Fuss-Catalan numbers} have their roots in
18th-century work on polygon dissections.  
Euler posed the problem of
counting the triangulations of a convex polygon, and Segner found a
recurrence in 1758.  
Thirty years later, Nikolaus Fuss 
 generalised   Euler's problem and von Segner's work  in a paper  \cite{Fuss1791} 
 presented at  the St. Petersburg academy. In 1838 
 Eug\`ene Catalan \cite{Catalan1838}   rediscovered  Euler's problem and gave it a 
  combinatorial interpretation   as the number  of sequences of legal (balanced) parentheses of given length. Since then the integers   solving Euler's problem  are referred to as Catalan numbers. The Catalan numbers form one of the most celebrated sequences in combinatorics, appearing in  many other problems as the
enumeration of Dyck paths, non-crossing
partitions, binary trees, and more than two hundred other combinatorial
structures~ \cite{Stanley_2011,book-Stanley}.  
The name
 Fuss-Catalan numbers  for the solution of the Euler-Fuss problem  was established in the 20th century.
  A  short historical overview  can be found in \cite{larcombe1998trail}. 

%
%
%

For integers $p \geq 2$ and $n \geq 0$, the \emph{Fuss-Catalan
number}, which we will denote by $\FC
_p  (n)$, is defined as\footnote{There is no unique established notation
for the Fuss-Catalan numbers.   
Another widely used notation is  related to
ours   by $A_n(p) = \FC_p (n)$. 
 } 
 \bee \la{FCdef} \FC
_p  (n)= \frac{1}{pn+1}\binom{ np+1}{n}. 
 \eee
The generating function of $ \FC ^{(p )}(n)$ satisfies the algebraic
equation
\bee \la{eqgenf} f = 1+z \,f ^p, \qquad f(z)= \sum_{n\ge 0}
\FC _p (n)\ z^n.  \eee
More generally, two-parameter Fuss-Catalan numbers or Raney numbers
are defined as \cite{Raney1960FunctionalCP}
 \bee \FC _{p,r}(n) = \frac{r}{np+r}\binom{np+r}{n}.
\la{defFC}
\eee
and for $r>1$, \bee \la{eqgenfr} f^{1/r}= 1+z \,f^{p/r}, \qquad f(z)=
\sum_{n\ge 0} \FC _{p,r}(n)\ z^n.  \eee
We will be concerned mainly with the case $r=1$.

The Fuss-Catalan number $\FC _p (n)$ admits two equivalent
combinatorial pictures.  Let us first recall the most familiar case
$p=2$ (Catalan).

\medskip

\emph{-- Tree picture (3-valent):} Planar binary trees with $n$
internal nodes; each internal node has $2$ children, with total graph
valency $3$.

\smallskip 

\emph{-- Chord diagram picture (2-valent).} Non-crossing pair
partitions of $\{1, 2, \ldots, 2n\}$ on a line (or circle): ways to
match the $2n$ points into $n$ pairs by chords drawn in the upper
half-plane (or disk) without crossings.

\medskip

The two descriptions are dual: the tree records the nesting structure
of the chord diagram.  We recall the exact match in Appendix
\ref{app:A}.  It is the second definition which can be naturally
extended to higher genus.  The two descriptions generalise for the
Fuss-Catalan numbers ($p \geq 2$) as follows.

\medskip
 
-- \emph{Tree picture ($(p+1)$-valent).} Plane $p$-ary trees with $n$
internal nodes; each internal node has $p$ children, total graph
valency $p+1$.

\smallskip -- \emph{$p$-hyperedge diagram picture ($p$-valent).}
Non-crossing partitions of $\{1, 2, \ldots, pn\}$ on a circle into $n$
blocks of size exactly $p$.  Each block is a $p$-uniform hyperedge
connecting $p$ points; the partition is non-crossing if no two blocks
``interlock" cyclically.
  One can  formulate this also as the number of ways to obtain
  a sphere by identifying groups of $p$ edges of a $np$-gon
  by  $p$-valent
hyperedges generalising 2-valent chords. 

%
%
 
 \medskip

 \subsection{Higher-Genus Catalan and Fuss-Catalan Numbers }

 \bigskip 

 \emph{ Higher genus Catalan numbers}

\smallskip

\noindent The genus $g$ Catalan numbers, which we denote by $
\FC _2^{(g)}(n)$, count the number of ways of obtaining a genus $g$
Riemann surface by identifying the sides of a $ 2n $-gon pairwise.
They have been first evaluated by Harer and Zagier in their famous
paper on the Euler characteristics of the moduli space of curves
\cite{HarerZagier1986}.  The Harer-Zagier numbers $\varepsilon _g(n )$
correspond in our notations to
\bee \varepsilon  _g(n)= \FC_2 ^{(g)}(n).  \eee

Harer and Zagier \cite{HarerZagier1986} reduced the computation of
Euler characteristics of the moduli space of smooth curves of genus
$g$ to the combinatorial problem formulated above, and derived an
explicit formula.  Itzykson and Zuber \cite{ItzyksonZuber1990}
simplified the derivation using the formulation of the numbers
$\varepsilon ^{(g)}(n)$ in terms of the expectation values in the Gaussian
$N\times N$ matrix ensemble with partition function
\bee \CZ_N\equiv \int d^N \X \ e^{- {1\over 2} \tr \X^2}, \eee
namely
\bee \sum_{g\le 2n} N^{1+n-2g}\ \varepsilon _g(n)= \<\tr \X^n \>_N
\equiv {1\over \CZ_N} \int d^N \X \ \tr \X^n\ e^{- {1\over 2} \tr
\X^2} .  \eee
The Harer-Zagier formula is thus equivalent to the expression for the
one-trace means~\cite{ HarerZagier1986}
\bee \la{HZP} { \<\tr \X^n \>_N\over (2n-1)!!}&= \hf \oint {dy\over
2\pi i} y^{-n-2} \({1+y\over 1-y}\)^N \\
&= \sum _{k=1}^{\min(N, n+1)} 2^{k-1} \binom{n}{k-1} \binom{ N}{k}
.
\eee
In this note we are concerned with the one-trace correlators.  The
formula for all genus multi-trace correlators in the Gaussian model
was first obtained in \cite{AlexandrovMironovMorozov2004}.

\bigskip 

\noindent \emph{ Higher genus Fuss-Catalan numbers}

\smallskip

\noindent There are at least two consistent definitions of the
higher-genus Fuss-Catalan numbers.  The first one comes from counting
the partitions of an ordered set of points.  It was studied recently
by J-B Zuber in \cite{Zuber2023}.  The second definition is a direct
generalisation of the original combinatorial problem.  Given a disk
with boundary divided into $np$ segments, the genus $g$ Fuss-Catalan
number, which we denote by $\FC_p ^{(g)}(n)$, counts the number of
ways to obtain a surface of genus $g$ by gluing groups of $p$ edges of
a $np$-gon by $p$-valent hyperedges.  The two definitions coincide
for $p=2$ (Catalan) but differ for $p>2$.  Below we describe the two
definitions in more detail.

\bigskip 

\emph{1.  Counting partitions by genus}

\smallskip

\noindent
We say that a partition $\a$ of the set  $\{1, \ldots,  n\}$ on a circle   
is of
 type $[\a]= [1^{\a_1}2^{\a_2}\cdots n^{\a_n}]$ if 
  it contains  $\a_1$ parts of length 1, $\a_2$ parts of length 2, etc,
 so that $\a_1+ 2 \a_2+\cdots= n$. 
 The total number of partitions of type $[\a]$  is 
 \bee C_{n, [\a]}= {n!\over \prod_{\ell=1}^n (\ell !)^{\a_\ell}\,  
 \a_\ell !  }\, .
 \eee

 A general partition $\a$ be described in terms of a pair of
permutations $\s$ and $\t$ where $\s$ is the cyclic permutation $(1,
2,..., n)$ and $\t$ belongs to the class $[\a]$.  
 It is postulated that

(i) the order of the elements is irrelevant 

\noindent 
and

 (ii) the relative position of parts is   irrelevant. 
 
\medskip

 \noindent 
By the assumption
(i), each cycle of $\t$ is an increasing list of integers.  The genus
g of the partition is then defined by \cite{Jacques1968}
\bee 
n+2-2g= \# (\text{cycles of }\t) + \# (\text{cycles of }\s) + \#
(\text{cycles of } \s\circ \t^{-1}) \eee
or with our specification for the permutation $\s$,
\bee
2g= n+1 - \sum_\ell \a_\ell -  \# (\text{cycles  of } \s\circ \t^{-1}) 
.
\la{genusper}
\eee
The total number of partitions  of type $[\a]$ 
is a sum over all genera $g$,
\bee
 C_{n, [\a]} = \sum  _{g\ge 0}C^{(g)} _{n, [\a]}. 
\eee

  Genus zero gives the standard
 non-crossing partition count.  In particular, the case of $p$-uniform
 partitions (all blocks of size exactly $p$) gives the Fuss-Catalan
 numbers. For many years, only  the enumeration of
non-crossing, or planar (genus 0)  partitions  was fully known
\cite{Kreweras1972}.  

 Recently, the generating function for the number of partitions of genus  
 $ 1$  and 2 were given in closed form  for $p=2,3$    in  \cite{Zuber2023,coquereaux2024}
  by reformulating the problem as counting ribbon graphs.  
%
%
%
%
%
%
%
%
%
As pointed out  by   Zuber~ \cite{Zuber2023},
the weighted sum
\bee
\la{weightedsum}
 C_{n, [\a]}(\xi) = \sum _{g\ge 0}C^{(g)} _{n, [\a]} \xi^g
\eee
can be formally defined as the Feynman graph expansion of a matrix integral by using 't Hooft
double-line notations  \cite{tHooft:1973alw}. 
  
Each partition defines a planar map\footnote{In this paper we 
consider planar maps and ribbon graphs as synonimes.} with $\a_\ell$ \ $\ell$-valent
vertices, for $\ell=1,2,...  , $ whose edges are numbered clockwise by
the elements of the partition, and a special $n$-valent vertex, with
its $n$ edges numbered anti-clockwise from $1$ to $n$.  Edges are
connected pairwise by matching their indices.  Two maps are regarded
as topologically equivalent if they encode the same partition.

\begin{figure}[t]
\centering 
\includegraphics[width= 14 cm]{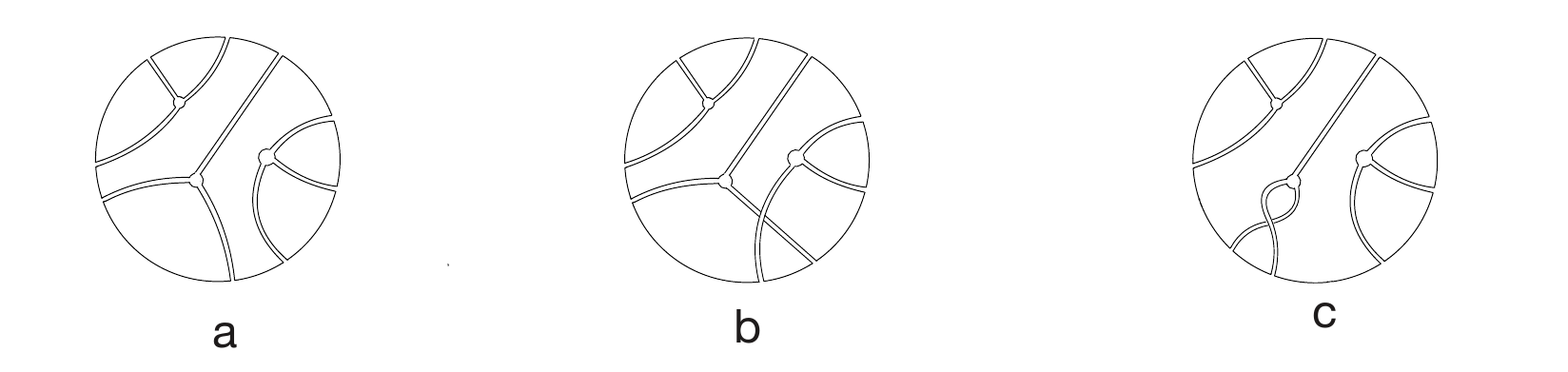}
\caption{\small   Examples of ribbon graphs  appearing in the evaluation of
\eqref{cummu} for $n=9$ and $[\a]=[3^3]$  of genus  $g=0$  (a) and  $g=1$  (b and c). The graph (c) is 
forbidden  by the {\em rc} rule  when  counting partitions but allowed when  counting  
planar maps.  
  }
\label{rbgfphs}
 \end{figure}

In addition, a {\em restricted crossing} (``$rc$'') 
from the condition (i)  is imposed:
 the edges connecting each $\ell$-vertex to the $n$-vertex
cannot cross one another, thus respecting their original cyclicity and
ordering.  Only crossings of edges emanating from {\em distinct} vertices
are allowed.  As an illustration we give some examples in fig. \ref{rbgfphs}.
The higher genus Fuss-Catalan numbers are given in this definition by
the particular case of a partition with $n$ cycles of length $p$:

The permutation $\s$ describes the connectivity of the $n$ points on
the circle, while $\t$ describes how these points are connected
through the other vertices.  The permutation $\s\circ \t^{-1}$
describes the circuits bounding clockwise the faces of the map, its
number of cycles giving the number $f$ of faces of the map.  Gluing a
disk to each face transforms the map into a closed Riemann surface, to
which we may apply Euler's formula
\bee 2-2g & = \# (\text{vertices}) - \# (\text{edges}) + \#
(\text{faces} ) \\
&= 1+ \sum_\ell \a_\ell -n + \# (\text{cycles of }
\s\circ \t^{-1}) \eee 
which reproduces \re{genusper}.


 The weighted sum \re{weightedsum} can be symbolically represented as  
 the expectation value in a Gaussian matrix model
 \cite{Zuber2023} 
 \bee
 \la{cummu} C_{n, [\a]}(1/N) =   \prod_\ell  
 N^{(1-\ell)\a_\ell -1}  \< :\tr \M^n :\
 :\prod _\ell { ( \tr \M^\ell /\ell)^{\a_\ell } \over \a_\ell !}
 :\>_{\! rc} 
 \eee
in a Gaussian $N\times N$ Hermitian matrix  ensemble, with $\xi = 1/N$.
The expectation value is computed by performing all allowed Gaussian contractions
\bee
\la{propxy}
  \langle \M_{ij}\,\M_{kl}\rangle \;=  \;     \delta_{il}\,\delta_{jk}.
\eee
The normal product $:\ :$ forbids Gaussian contractions between the 
matrices inside  the product; here it forces all edges to reach the
$n$-vertex. The  allowed contractions are further restricted by the {\em rc} condition
which states that no crossings are allowed for the lines emanated by the same vertex.

The expectation value evaluates the sum over all    ribbon graphs
with the above restrictions. 
 For such graphs the normalisation factor 
on the rhs of \eqref{cummu}  can be represented as   $N^{ \# (\text{vertices}) - \# (\text{edges}) }$
 which,  combined with the factor $N^{\#\text{faces}}$ from the sum over 
  indices, produces the topological factor $N^{-2g}$.

 The   genus $g$ Fuss-Catalan numbers in the sense of partitions
 are given   by the $ (1/N)^{2g}$ term of  \eqref{cummu} for 
  a partition with $n$ cycles of length $p$:
 \bee
  \FC_p ^{(g)} (n) \Big|_{\text{by partitions}} \ =\  [N^{-2g}] C _{np , [p^n]} (1/N).
  \la{FCparts}
  \eee
  %
 %

 Unfortunately  there is no simple way to implement the {\em rc} condition from the rule (i) in  the matrix  integration measure. 
  The matrix model \eqref{cummu}    is thus only  a formal tool to generate the right combinatorics for the sum over ribbon graphs. In general, the
enumeration of the partitions according to their genus is still an
open problem.


 \bigskip 

 \emph{2.  Counting planar maps by genus }  

\smallskip

\noindent The second definition is more natural from the perspective
of the `t Hooft topological expansion of matrix integrals and in view
of the relation to moduli spaces of curves.  In this definition, the
restricted crossing condition in the matrix integral \re{cummu} is
abandoned and we have access to the full power of the matrix integral
formulation.  The sum over ribbon graphs  then represents a connected
multi-trace correlator in the standrd   Gaussian matrix model,
 \bee 
 \la{cummu2} \CC_{n, [\a]}(1/N) =  
 N^{(1-\ell)\a_\ell -1}  
  \prod_\ell {
 N^{(1-\ell)\a_\ell }\over \ell^{\a_\ell} \a_\ell !} \< \tr \X^n \prod
 _{\ell=1}^n {( \tr \X^\ell /\ell)^{\a_\ell }\over \a_\ell!} \> _{\!\!\!\text{conn}} .
 \eee

Once all the correlation functions in the Gaussian model are known,
the higher genus Fuss-Catalan numbers can be obtained directly from
the above formula by  extracting the coefficient of $(1/N)^{2g}$.  
Explicit formulas for all genus $k$-trace connected correlators in the Gaussian model 
have been obtained  with $k=1$ \cite{HarerZagier1986},
$k=2$ \cite{Morozov_2009} and $k=3$ \cite{A.Morozov:2010ab}.
The $k\ge 4$ correlators have been
obtained in terms of matrix resolvents and hypergeometric functions in
\cite{Dubrovin_2017}.  All correlators are computable in principle by
topological recursion \cite{eynard2016counting}, but practically the recursion procedure becomes  
inefficient at higher orders. This is why we will follow here a different approach, which exploits
the fact that the problem can be reduced to the computation of
determinants.

%
%

 \section{ The Fuss-Catalan  matrix model}
   
 Consider the ensemble of two  Hermitian $N \times N$ matrix variables
   $X, Y \in \HN$  with partition function
   \bee
   \la{defpf}
   \CZ_{N,p} = \int d\X d\Y\ e^{-  S(\X,\Y)}
   \eee
and action\footnote{ The matrix model \re{defpf}-\re{eq:action} is
the poor cousin of the family of the generalised Kontsevich models
(GKM) \cite{Kharchev:1991cu}.  Closely related models have been
studied by Br\'ezin and Hikami
\cite{Br_zin_2007,Br_zin_2008,Br_zin_2009}.  The main difference with
the GKM models is that our model generates planar maps with $p$-valent
vertices while the GKM generates trees with $(p+1)$-valent vertices.
Our model gives the "hyperedge-picture" generalisation while the
ciliated maps from the GKM model
\cite{belliard2023topologicalrecursiongeneralisedkontsevich} are
closer to the combinatorial definition of the classical Fuss-Catalan
numbers in terms of $(p+1)$-valent trees.  }
\begin{equation}
\label{eq:action}
  S(\X, \Y) \;=\; \tr(\X\Y) \;-\;   {\tr(\Y^p)\over p},
  \qquad p \geq 2.
\end{equation}
 Since we are interested by the combinatorial side only, we will not
 discuss the issue of convergence and the choice of the contours of
 integration.

 We claim that the genus-$g$ Fuss-Catalan numbers defined in the
 Introduction as genus-$g$ planar maps can be extracted from the $1/N$
 expansion of the one-trace expectation values of the moments of the
 variable $\X$, namely
\bee
  \sum_{g\ge 0}  \FC^{(g)}_p (n) \ N^{-2g} &= 
  {1\over N  ^{1+(p-1)n }}
  \< \tr  (\X^{pn})\>_{N,p}
  \la{Cng}
 \eee
 \bee 
 \< \tr (\X^{pn})\>_{N,p} 
 & \equiv {1\over \CZ_{N,p} } \int d\X\ d\Y \ \tr (X^{pn})\ e^{ -
 N\tr \( \X\Y -{ \Y^p\over p} \) } .  \la{expvFCMM} 
 \eee    

To see that, consider the formal Feynman expansion of \re{Cng}
treating the second term in the action \re{eq:action} as a
perturbation around the ``Gaussian'' action $\tr \X\Y$.  The Feynman
rules involve
  
  -- $\Y$-vertex from $ \tr(\Y^p)/p$.  This is a $p$-valent
  hyper-vertex (one half-edge for each $\Y$-factor in the trace
  $\tr(\Y^p)$,
\bee
 {1\over p} \tr(\Y^p)\ \ \to\ \ {1\over p} 
\d_{i_1,j_2}\d_{i_2, j_3}...\d_{i_p, j_1}
\la{XYprop}
\eee
 -- $\X$-cycle from $ \tr(\X^{pn})$
 \bee \tr(\X^{pn})\ \ \ \to\ \ \ {1\over N} \d_{i_1,j_2}\d_{i_2,
 j_3}...\d_{i_{np}, j_1} 
 \eee 
 -- $\X\Y$-``propagator'' from $
 \tr(\X\Y)$.  This is 2-valent edge connecting one half-edge of $X$ to
 one half-edge of $Y$:
\bee
\la{propxyxy}
  \langle \X_{ij}\,\Y_{kl}\rangle \;=  \;     \delta_{il}\,\delta_{jk}.
\eee

Performing all $np$ contractions \re{XYprop} between the $\X$-cycle
and $n$ $\Y$-vertices, one obtains a sum of all possible $p$-uniform
hyperedge diagrams on the $\X$-cycle : each $\Y$-vertex is a
$p$-valent hyperedge, connecting $p$ points on the cycle.  This is
exactly the $p$-valent hyperedge picture for the higher genus
Fuss-Catalans, generalising the chord-diagram picture of $p=2$.

A diagram which appears
as a result of Wick contractions represents a ribbon graph with double
lines and characterised by $n$ $p$-valent vertices, $pn$ double-line
propagators and $F$ faces spanning the ``index loops''.  With the
factor $1/p$ in the action \re{eq:action}, the contribution of a graph
is just a power of $N$.  By the Euler relation, this power is
 \bee
 {1\over N^{n(p-1)+1} }\ 
 N^{   F }=  N^{(\text{vertices)}-(\text{edges) +(\text{faces})} }=N^{ -2g}. 
 \eee
Hence the expectation value $ \<\tr \X^{np}\>/N^{n(p-1)+1} $ indeed
gives the sum on the lhs of \re{Cng},   with  the genus $g$ Fuss-Catalan
number  defined as the number of ribbon graphs of genus $g$ having
$n$ vertices and $pn$ lines, as illustrated in fig.  \ref{rbgfphs}.
   
   Concerning the  definition of higher-genus Fuss-Catalan numbers
   by counting partitions,
   a version of the FC matrix model whose correlators obey the  \emph{rc} restriction is 
    formulated in Appendix \ref{appE}.
 
\bigskip

\noindent
\emph{Spectral curve}

\medskip

\noindent The spectral curve can be derived proceeding e.g. as in
section 2 of \cite{Kazakov:2004du}.  Since the integrand is $U(N)$
invariant, we can transform the matrix integral into an integral over
the eigenvalues $\{ x_i\}$ and $\{y_i\}$.  Applying the
Harish-Chandra-Itzykson-Zuber formula, we write the matrix integral
\re{defpf} as
\bee
\CZ_{N,p} =[\text{volume of }U(N)]\times \int \prod_{j=1}^N d x
 _j d y _j \ \Delta(x)\Delta(y) e^{ \sum _i (-x_i y_i + {1\over p}
 y_i^p)} \la{eigenvi} 
 \eee
 where
 \bee
 \Delta(z) = \prod_{j<k} (z_j-z_k) 
 \eee
is the Vandermonde determinant.  At large $N$ the integral is
saturated by a saddle point.  This gives conditions on the effective
action $S_{\text{eff}}(x,y)$ for a pair of probe eigenvalues
\bee
S_{\text{eff}}(x,y)
&= \Omega(x) +\tilde\Omega(y)  + xy  - {1\over p} y^p ,
 \\
 \Omega (x)&=  -   \sum_{j} \log  (x-x_j)
\\
 \tilde\Omega(y)&=   -     \sum_{j}  \log (y- y_j) .
\eee
The  saddle point equations read 
\bee
x= y^{p-1}-\tilde \Omega '(y) \equiv X(y),\qquad y=- \Omega'(x)
\equiv Y(x)
\eee
where the functions $X$ and $Y$ are inverse to each other if
considered as {multivalued} meromorphic functions defined on their
Riemann surfaces.  On the physical sheets they satisfy (again on the
physical sheets) the asymptotic relations
\bee x=X(y)&= y^{p-1} + {1\over y} + o(1/y^2), \qquad y\to\infty \cr
y=Y(x)&=   {1\over x} + o(1/x^2), \qquad  \qquad  \quad  x\to\infty.
\eee
These conditions and the fact that the two meromorphic functions are
inverse to each other and have no other poles except those at infinity
determine them completely.  The functions $x=X(y)$ and $y=Y(x)$ are
determined by the algebraic equation (the spectral curve)
\bee
\la{SCFC}
 xy = 1+ y^p
\eee
which encodes the $g = 0$ data of the Fuss-Catalan problem.  Indeed,
setting $z=x^{-p}$ and $f= x\,y$ in the generating-function equation
\re{eqgenf}, $f=1+z f^p$, gives $x y = 1 + x^{-p}(x y)^p = 1+ y^p$,
which is precisely \re{SCFC}.  Once the
spectral curve is known, one can obtain in principle all higher-genus
coefficients by doing topological recursion.

\section{Exact identities (sum rules) from the matrix model}

We will obtain exact sum rules for the higher genus Fuss-Catalan
coefficients by evaluating explicitly the expectation values on the
rhs of \re{Cng}.  To evaluate the expectation value of $\tr \X^{np}$,
we use Br\'ezin-Hikami contour integral representation
\cite{Br_zin_2007,Br_zin_2008,A.Morozov:2010ab}.  Define the
exponential 1-point density as
\bee
\label{eq:e-def}
e_p(s\mid N) \: :=\;
\bigl\langle\operatorname{tr}\,e^{s\X}\bigr\rangle_{N,p} 
\;=\;\sum_{k\geq
0}\frac{s^k}{k!}\,\bigl\langle\operatorname{tr}(X^k)\bigr\rangle_{N,p} 
.
\eee
By the $\mathbb{Z}_p$ discrete symmetry
$\langle\operatorname{tr}(\X^k)\rangle = 0$ unless $p\mid k$, and
$e_p(s\mid N)$ is a power series in $s^p$.
 
For the Gaussian measure, $p=2$, the rhs of \re{eq:e-def} can be
represented as a single contour integral by the Br\'ezin-Hikami GUE
formula \cite{Br_zin_2008,A.Morozov:2010ab},
  \bee 
  \la{BH2} e_2(s\mid N) \;=\;\frac{e^{s^2/2}}{s}\oint\frac{\dd
  u}{2\pi i}\,e^{us} \Bigl(1+\tfrac{s}{u}\Bigr)^N\qquad (p=2).  
  \eee
The contour encircles $u=0$ counterclockwise; the integrand has a pole
of order $N$ at $u=0$ from the factor $(1+s/u)^N$, and the exponential
is entire function of $u$ (and of $s$).


The $p\ge 2$ generalisation of Br\'ezin-Hikami formula, the
derivation of which we give in Appendix \ref{BHproof}, reads
\begin{equation}
\label{eq:BH}
  e_p(s\mid N)
  \;=\;
  \frac{1}{s}\oint_{u=0}\frac{\dd u}{2\pi i}\;
  \exp\!\Bigl[\tfrac{1}{p}\bigl((u+s)^p - u^p\bigr)\Bigr]\,
  \Bigl(1 + \frac{s}{u}\Bigr)^{N}.
\end{equation}
After the substitution $u = sw$ and series expansion in $q = s^p$,
this gives\footnote{Eq. \re{eq:trYpn} is a particular case of an
 integral formula proved by A. Hock, eq. (3.2) of 
\cite{Hock2024}.  We thank  J-B Zuber for communicating to us Hock's paper.} 
\bee
\label{eq:trYpn}
  \langle\operatorname{tr}(\X^{pn})\rangle_{N,p} 
 & \;=\; \frac{(pn)!}{p^n n!}\,\bigl[w^{N-1}\bigr]\,
  \bigl((w+1)^p - w^p\bigr)^n\,(w+1)^N
  \\
  &\;= \; \; \frac{(pn)!}{p^n n!}\, \oint _\infty 
  {\dd w\over 2\pi i}
  \bigl((w+1)^p - w^p\bigr)^n\, \( 1+{1\over w}\)^N. 
  \eee
Changing the variable as $w = 1/t$, we obtain the following sum rule
for the genus-$g$ Fuss-Catalan numbers
 \bee
\label{eq:yoursumrule}
\sum_{g=0}^{g_{\max}} \FC^{(g)}_p (n)  \;N^{-2g}
   & \equiv  N^{-1 -(p-1)n}
    \langle\operatorname{tr}(\X^{pn})\rangle_{N,p} 
   \\
  & \;=\; \frac{(pn)!}{p^n\,n!  N^{(p-1)n+1}}\; \oint_0 {\dd t\over
  2\pi i} t^{- pn-2} \,\bigl((1+t)^p-1\bigr)^n(1+t)^N. 
  \eee
We may rewrite~\re{eq:yoursumrule} in a more transparent form by
expanding
\begin{equation}
  \bigl((1+t)^p - 1\bigr)^n \;=\; \sum_{r=0}^{p n } T_p ^{(n
  )}(r)\;t^r\, , \qquad T_p ^{(n )}(r) \;=\; \sum_{k=0}^{n }(-1)^{n
  -k}\binom{n }{k}\binom{p k}{r}.
  \label{eq:GF}
\end{equation} 
 We find, and this is our main result\footnote{Another 
 explicit formula, equivalent  to \eqref{eq:exact2},  and which can be deduced from Zagier  \cite{zagier1995distribution} or Stanley  \cite{stanley2009enumerativeresultscyclespermutations}, was given explicitly by Huang and Yang in  \cite{huang2025enumerationlhypermaps}. The formula derived in \cite{huang2025enumerationlhypermaps} contains a single sum of products of two binomial coeﬃcients.  Also,  \cite{huang2025enumerationlhypermaps} presents  a residue formula equivalent to \eqref{eq:yoursumrule}.
We thank one of  the referees for bringing  these works  to our knowledge.}

\bee
    \sum_{g=0}^{g_{\max}} 
   \FC^{(g)}_p (n)  \;N^{-2g}
  \;=\;
  \frac{(p n )!  }{p ^n \,n !\;N^{(p -1)n +1}}
  \sum_{r=n }^{p n } T_p ^{(n )}(r)\;\binom{N}{p n +1-r}
  \label{eq:exact2}
\eee
 Equivalently, we can write \re{eq:yoursumrule}, multiplying by $N^{(p-1)n+1}$,
 as
\begin{equation}
\label{eq:sumrule-poly}
  \sum_{g=0}^{g_{\max}} \FC^{(g)}_p (n) \,N^{(p-1)n+1-2g}
  \;=\;
  \frac{(pn)!}{p^n\,n!}\sum_{s=1}^{(p-1)n+1} \binom{N}{s}
    \,
  [x^{pn+1-s}]\bigl((1+x)^p-1\bigr)^n
\end{equation}
which expresses the genus-expansion polynomial in $N$ explicitly
as a finite linear combination of binomials $\binom{N}{s}$ with
coefficients that are alternating binomial sums.
At $p = 2$, the sum rule \eqref{eq:sumrule-poly}  
reproduces the   Harer-Zagier polynomial~\cite{HarerZagier1986,Pittel2016},
eq. \re{HZP}.
    
For $N=1$ the sum rule becomes particularly simple:
\bee
\label{eq:N1}
  \sum_{g=0}^{g_{\max}} \FC^{(g)}_p (n) \;=\; \frac{(pn)!}{p^n\,n!}
\eee
with the rhs being the number of partitions of $\{1,\ldots,pn\}$ into
$n$ blocks of size $p$.

 \section{ Genus  $g$ Fuss-Catalan numbers 
 }
 \label{sec:master}

\subsection{General formula}

The sum rule we derived can be analytically continued in $N$ and
expanded in the negative powers of $1/N^2$.  Using this property we
can extract the FC numbers of given genus $g$.

It is convenient and useful for the applications to compute the ratios
\bee
\la{defRg}
  \K^{(g)}_p (n) \;:=\; \frac{ \FC_p^{(g)}(n)}{ \FC_p^{(0)}(n)}
  \qquad ( p \geq 2 ,\ n \geq 1  ,\  g \geq 1).
  \eee
 We find that the ratio is a polynomial in $n$  and  $p$ 
 determined by 
 \bee
\label{eq:master}
   \K^{(g)}_p (n) \;=\;  { (pn-n+1)!\over (pn -n +1-2g)!}
\bigl[t^{2g}\bigr]\,G_n(t) 
\eee
where the second factor is the coefficient of the power $t^{2g}$
in the series expansion of the function 
\bee
  G_n(t) \;:=\;  e^t\,\frac{[h(p t)]^n}{[h(t)]^{pn+2}}, 
  \qquad 
  h(t) := \frac{e^{t}-1}{t} .
\eee
 at $t=0$.

\bigskip

\noindent
 \emph{Proof}:
Apply the substitution $w = 1/(e^t-1)$ in~\eqref{eq:trYpn}.  
 The contour integral becomes 
 \bee
  {1\over 2\pi i}\oint\limits _0 {d t  \, e^t\over  (e^t-1)^2}\
 e^{t N} \left(e^{p t} \left(e^t-1\right)^{-p}-\left(e^t-1\right)^{-p}\right)^n
   \;=\;  \underset{t=0}{\mathrm{Res}} \Bigl[\tfrac{e^{(N+1)t}\,(e^{pt}-1)^n}{(e^t-1)^{pn+2}}\Bigr] 
\eee
and the residue is the coefficient of $t^{(p-1)n+1}$ in $p^n
e^{(N+1)t} [h(pt)]^n / h(t)^{pn+2}$.  Hence
\bee
  \langle\operatorname{tr}(\X^{pn})\rangle _{N,p} 
  \;=\; \frac{(pn)!}{n!}\,\bigl[t^{(p-1)n+1}\bigr]\,
  \frac{e^{(N+1)t}\,[h(pt)]^n}{[h(t)]^{pn+2}}.
\eee
After dividing  by $N^{(p-1)n+1} \FC^{(0)}_p (n) $ 
we   express the rhs as a power sum in $1/N^2$:  
\begin{align*}
  \frac{ \sum_{g=0}^{g_{\max}} 
   \;N^{-2g}\ \FC^{(g)}_p (n)  
}{  N^{(p-1)n+1} \,\FC^{(0)}_p (n) } 
  &= \frac{((p-1)n+1)!}{N^{(p-1)n+1}}\,\bigl[t^{(p-1)n+1}\bigr]\,e^{Nt}\,G_n(t) \\
  &= \frac{((p-1)n+1)!}{N^{(p-1)n+1}}\,\sum_{m=0}^{(p-1)n+1}
  \frac{N^{(p-1)n+1-m}}{((p-1)n+1-m)!}\,\bigl[t^m\bigr]\,G_n(t) \\
  &= \sum_{m=0}^{(p-1)n+1}\frac{((p-1)n+1)!}{((p-1)n+1-m)!}\,
  \bigl[t^m\bigr]\,G_n(t)\,N^{-m}.
\end{align*}
 Reading coefficient $[N^{-2g}]$ (taking $m = 2g$) we get 
 \bee
 \K^{(g)}_p (n) \;=\;
 \frac{((p-1)n+1)!}{((p-1)n+1-2g)!}\,\bigl[t^{2g}\bigr]\,G_n(t) \;=\;
 \bigl[(p-1)n+1\bigr]_{2g}\,\bigl[t^{2g}\bigr]\,G_n(t) \ \  \square\ \ \ 
 \eee

\subsection{ The ratio $ \K^{(g)}_p= \FC^{(g)}_p /\FC _p^{(0)} $ as a polynomial in
$n$}

The master formula~\eqref{eq:master} extracts {all} coefficients of $
\K^{(g)}_p (n) $ as polynomials in $n$,

\bee
\label{eq:structure}
  \K^{(g)}_p (n)
   \;=\;
    \sum_{k=1}^{3g} (p-1)^k\,Q_ k  \,n^k
   ,
\eee
where the coefficients $Q_k$ being polynomials in $p$ of degree $\le
2g-1$.  At higher orders $[n^k]$ for $k\geq 2$, the structure becomes
richer; the $p$-dependence is no longer purely $(p-1)$.

The master formula~\eqref{eq:master} expresses the ratio $
\K^{(g)}_p (n) $ as a polynomial in $n$.  The highest and the lowest
coefficients of this polynomial are relatively easy to compute.
 
 \medskip
 
\noindent -- the leading $n^{3g}$ coefficient is 
 for all $g \geq 1$ and $p \geq 2$,
\bee
\label{eq:leading}
  Q_{3g} (p)  \;=\;  {1\over g!} \({p\over 24}\)^g   
 .
\eee
-- The linear term in $n$ is a bit less trivial.  The computation is
sketched in Appendix \ref{sec:n1-factor}.  The result is
 \bee \la{Bernid}
 \text{Bernoulli identity:} \hskip 2cm 
 Q_1(p) \;=\; - \,\frac{B_{2g}}{2g}.\hskip 3cm  
\eee 

\subsection{Closed form for $g\le 3$}

\noindent
\subsection*{$g=1$}
\vskip -0.8cm
\bee
 \K_p  ^{(1)}  (n)  
&  \;=\; \sum_{k=1}^{3} (p-1)^k\,Q_k \,n^k,
  \\
 Q_1 & \;=\;  -  \tfrac{1}{12}=  -\tfrac{B_2}{2},
 \\
  Q_2 & \;=\;   \tfrac{p-2}{24}
\\
Q_3  &\;=\;   \tfrac{p}{24}.
\eee

\subsection*{$g=2$}
\vskip -0.8cm

\bee
\label{eq:R2}
  \K_p   ^{(2)}  (n) &\;=\; \sum_{k=1}^{6} (p-1)^k\,Q_k\,n^k,
  \eee
  where the coefficient polynomials are
\bee
\label{eq:Qktop}
  Q_1 &= \tfrac{1}{120} = -\tfrac{B_4}{4},\\
  Q_2  &=-\tfrac{p^3+p^2+11 p+6}{1440} ,\\
  Q_3  &=   \tfrac{ p^3 + 6\,p^2 + 11\,p - 24}{2880},\\
  Q_4 &= \tfrac{ 4p^3 - p^2 + 44\,p + 24}{5760},\\
  Q_5  &= -
   \tfrac{ \,p\,(p^2 + 6\,p + 11)}{2880},\\
  Q_6 &= 
    \tfrac{1}{2!} \, \(\tfrac{ p}{24}  \)^2.
\eee

\subsection*{$g=3$}
\vskip -0.8cm
\bee
\label{eq:Rg3}
  \K_p ^{(3)}(n)  &\;=\; \sum_{k=1}^{9}  (p-1)^k
   \,Q_k \,n^k,
 \\ 
  Q_1 &\;=\; -\tfrac{ 1}{252}= -\tfrac{B_6}{6},\\
	Q_2&\;=\; \tfrac{ 4 p^5+4 p^4+25 p^3+25 p^2+151 p+130 }{30240},\\
	  Q_3&\;=\; - \tfrac{ 52 p^5+178 p^4+451 p^3+1081 p^2+1963 p-900
	  }{362880},\\
	   Q_4&\;=\; - \tfrac{ 20 p^5-71 p^4-36 p^3-421 p^2+755
	   p+1000}{241920},\\
	   Q_5&\;=\; \tfrac{ 40 p^5+103 p^4+222 p^3+628 p^2+1510 p+432
	   }{290304},\\ Q_6&\;=\; - \tfrac{ 48 p^5+398 p^4+825 p^3+2400
	   p^2+1812 p+160}{967680},\\
        Q_7&\;=\; 
        \tfrac{ 16 p^4+394 p^3+1353 p^2+2368 p+604 }{2903040}
        ,\\ Q_8&\;=\;    - \tfrac{  2 p^2+17 p+12 }{138240} ,\\
           Q_9&\;=\;   \tfrac{1}{3!} \, \(\tfrac{ p}{24}  \)^3
\eee

\subsection{Two-trace correlators at leading order}

In principle, one can use the Br\'ezin-Hikami method \cite
{Br_zin_2008} to compute the multi-point correlators in the FC matrix
model by considering the connected expectation value of a product of
exponential operators generalising \re{eq:e-def}.  The computation
goes out of the scope of the present short note, but let us mention
here the expression for the genus-zero two-point FC numbers which
turns out to be particularly simple.

The two-trace correlator  expands as
\begin{equation}
\label{eq:genus-exp}
  \bigl\langle \tr(\X^{p n_1})\,\tr(\X^{p n_2})\bigr\rangle
  _{N,p} \Big|_{\text{connected}} \;=\; N^{(p-1)(n_1+n_2)}\sum_{g \ge
  0}\,\FC_p ^{(g)}(n_1, n_2)\,N^{-2g}
\end{equation}
with the $g=0$ term given by 
\bee \label{eq:c-zero}
  \FC_p ^{(0)}(n_1, n_2) 
 & \;=\; (p-1)\,\frac{n_1\,n_2}{n_1 + n_2}\,
  \binom{p\,n_1}{n_1}\,\binom{p\,n_2}{n_2}.
\eee

\section{Conclusion}

In this note we studied a higher-genus generalisation of the
Fuss-Catalan numbers.  For $p>2$, there are at least two consistent
definitions of genus $g$ FC numbers.  We focused on the one which is
more natural from the string theory perspective.  We constructed a
two-matrix model, the Fuss-Catalan matrix model, eqs.
\re{defpf}-\re{eq:action}, which generates the $g\ge 1$ Fuss-Catalan
numbers $\FC_p ^{(g)}(n) $.  At $p = 2$, $\FC_2 ^{(g)}(n) =
\varepsilon^{(g)}(n)$ (Harer-Zagier numbers) and $\FC^{(0)}_2(n) = C_n$
(Catalan numbers).
 
The Fuss-Catalan matrix model provides one more example for the
realisation of topological gravity via matrix models first suggested
by Witten \cite{Witten1991,Witten:1993mgm}.  We determined the
one-trace expectation values in the FC matrix model for any $g$ from a
set of exact sum rules.  Our result is a $p>2$ generalisation of the
Harer-Zagier formula \cite{HarerZagier1986}.

We find that the genus-$g$ FC numbers normalised by the classical
($g=0$) FC numbers, \re{defRg}, are polynomials of degree $3g$ in $n$,
with vanishing constant term for $g \geq 1$,
\bee 
\la{polyQ}
  \K_p ^{(g)}(n) \;:=\; \frac{\FC_p ^{(g)}(n)}{\FC_p ^{(0)}(n)}
    \;=\;
    \sum_{k=1}^{3g} (p-1)^k\,Q_ k  \,n^k.
\eee 
where $Q_k$ are polynomials in $p$. 
The top and the bottom coefficients of the polynomial $ \K_p ^{(g)}(n) $
are obtained in a closed form; they are supposed to reflect distinct geometric
invariants of the moduli space $\Mbar_{g,n}^{1/p}$ of $p$-spin curves.
The intermediate coefficients interpolate via Hodge integrals.

-- The top coefficient  gives  the one-point $p$-spin
Witten-Kontsevich intersection number on $\Mbar_{g,1}^{1/p}$.  In the
normalisation of~\cite{LiuVakilXu2017,Br_zin_2008} one has
$\langle\tau_{3g-2,0}\rangle^{(g)}_{p\text{-spin}}=(p-1)^g/(24^g g!)$
(for $g=1$ this is the Liu-Vakil-Xu value $(p-1)/24$), so that
\bee
\label{eq:top}
 [n^{3g}]\,\K^{(g)}_p (n) & \;=\;(p-1)^{3g}\,Q_{3g}\\
 &  \;=\;  (p-1)^{3g}\; \frac{p^g}{24^g\,g!}
   \;=\; p^g\,(p-1)^{2g}\,
   \langle\tau_{3g-2,0}\rangle^{(g)}_{p\text{-spin}}  .
\eee
The factor  $p^g(p-1)^{2g}$ 
 is the matrix-model normalisation relating the FC ribbon-graph count to the one-point Witten-class integral. At  $p=2$ the  piece $(p-1)^{2g}$  
 equals one and the normalisation reduces to the standard Gaussian factor 
 $2^g$. For $p\ge 3$ the additional $(p-1)^{2g}$ is a genuine $p$-spin contribution.

 -- The bottom coefficient  is  the 
 orbifold Euler characteristic of $\Mbar^{1/p}_{g,1}$:
\bee
\label{eq:bottom}
[n^{1}]\,\K^{(g)}_p (n)\; =\;   (p-1)\; Q_1\;=\; - \; (p-1) \,\frac{  B_{2g}}{2g}
   \;=\;  \chi(\Mbar_{g,1}^{1/p})\; .
\eee
The $(p-1)$ factor reflects the orbifold structure of the
$p$-spin cover $\Mbar_{g,1}^{1/p} \to \Mbar_{g,1}$.  At $p = 2$, this
reduces to the Harer-Zagier/Penner formula
$\chi(\Mbar_{g,1}) = -B_{2g}/(2g)$~\cite{HarerZagier1986, Penner1988}.

Furthermore, the intermediate coefficients of the polynomial
\re{polyQ} are expected to encode $p$-spin Hodge integrals.  We will
postpone their exact interpretation as well as the analysis of the
multi-trace correlators for future work.

 The polynomial form of the normalised FC numbers  $N_p^{(g)}(n)$
 reflects the large-$n$ behaviour of the trace correlators, which is controlled by the edge 
 behaviour of the spectral density of the FC matrix model. For $p=2$ this is the well-known Wigner semicircle with its Airy edge; 
 for general  $p$  the spectral density of 
 $\X$  in the FC matrix model is  the Fuss–Catalan distribution, whose edge singularity is again of Airy type. 
 Indeed the $n^{-3/2}$ prefactor 
  in the large-$n$ 
 asymptotics of $C^{(0)}_p(n)$ signals a square-root branch cut in the resolvent, i.e. a two-sheeted spectral curve whose sheets meet at the endpoints of the support — the generic Airy geometry.
 
 Okounkov and Pandharipande  \cite{okounkov2001}  observed that at 
$p=2$ the double-scaling limit of the Gaussian ensemble at the edge yields the Witten–Kontsevich intersection numbers directly, bypassing Kontsevich's Airy matrix model. Our master formula \eqref{eq:master} is the analogous statement for $p\ge 2$;
the leading $n^{3g}$ coefficient of $N_p^{(g)}(n)$
is precisely the one-point  $p$ -spin intersection number, and the FC matrix model plays the same role for the
$p$-spin intersection theory that the Gaussian ensemble plays for the classical Witten–Kontsevich case.

\bigskip

\subsection*{ Acknowledgements}

I.K.\ thanks Andrey Okounkov, Mauricio Romo, and Jean-Bernard Zuber
for discussions.
The work of A.N.K.\ was supported by the Beijing National Science
Foundation (IS24006).

\bigskip

 \appendix

 \section{ Relation between the two combinatorial definitions of FC}
\la{app:A}

\begin{figure}[h!]
        \centering
\begin{minipage}[t]{0.9\linewidth}
            \centering
                         \includegraphics[width= 9 cm]{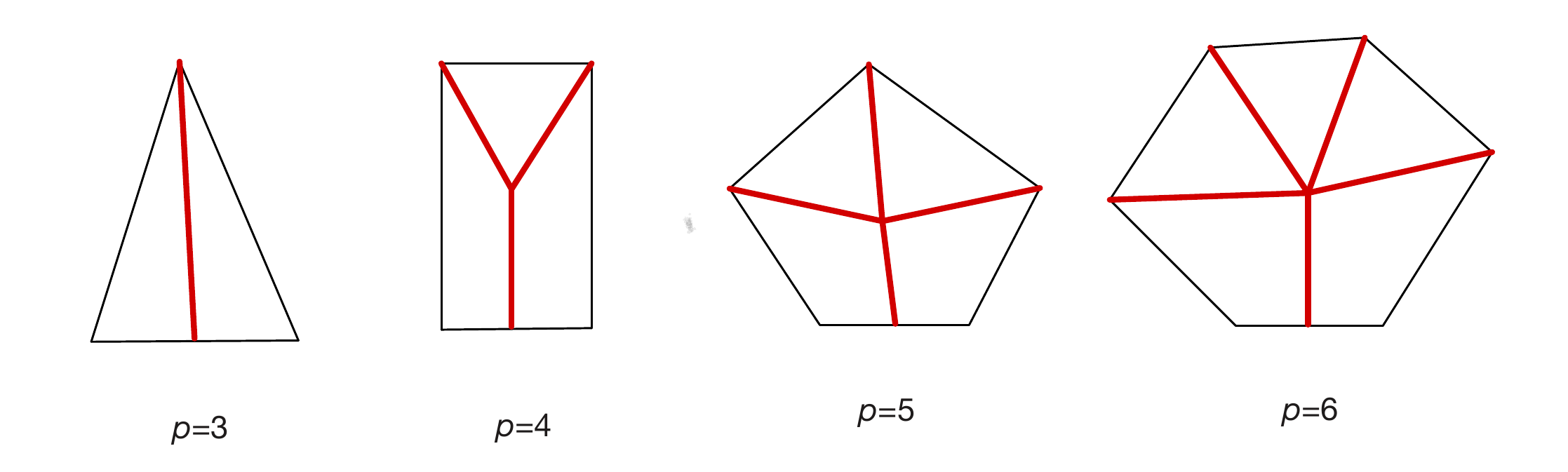}
  \caption{\small    Correspondence between $(p+1)$-gons and $p$-vertices.}
     \label{fig:Polygons}
          \end{minipage} 
\end{figure}

Let us first note that with  any $(p+1)$-gon  with a marked side (root)  we can 
assign  a $p$-valent vertex so that one of the edges ends at the root while 
the remaining $p-1$ edges end at the corners non-adjacent to the root (fig. \ref{fig:Polygons}). This correspondence  will  be used  below to
associate a planar $p$-hyperedge diegram with each $(p+1)$-valent tree. 

\begin{figure}[h!]
        \centering
\begin{minipage}[t]{0.9\linewidth}
            \centering
                         \includegraphics[width= 13 cm]{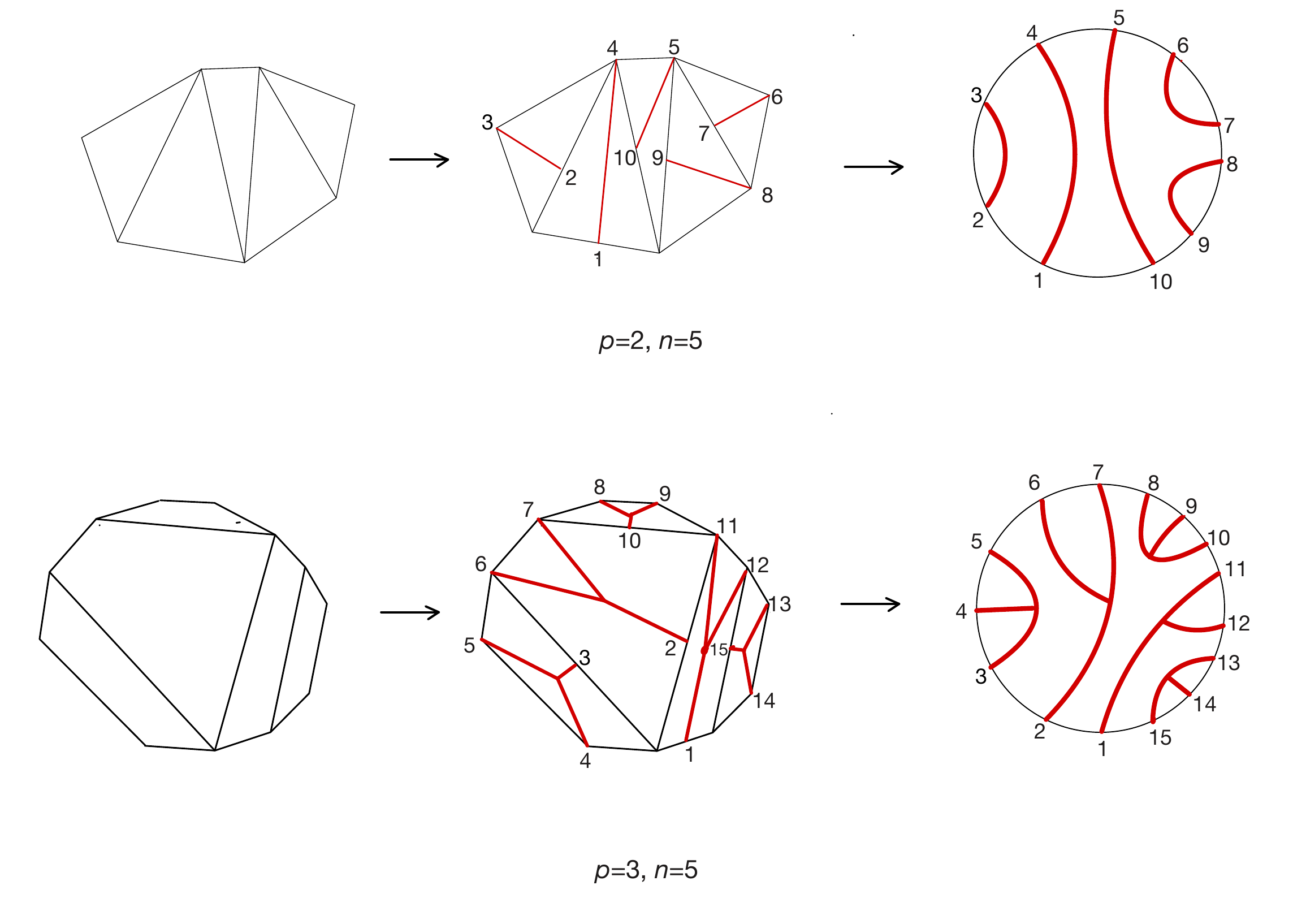}
  \caption{\small Two examples of the correspondence between dissections of a
  $(n(p-1)+2)$-gon into $n$ $(p+1)$-gons and planar graphs with $n$
  vertices of valence $p$ and $pn$ external lines.
  }
     \label{fig:dissectionplanar}
          \end{minipage} 
\end{figure}

Consider a dissection of a $((p-1)n+2)$-gon into $n$ $(p+1)$-gons.
 Choose a root   and  decorate the $(p+1)$-gon
sharing this root by a $p$-valent vertex.  Label the edge  which  ends 
at the root by 1.  If the leftmost from the root side of the  $(p+1)$-gon 
is also a side of the  original  polygon,  label the
edge associated with the corner not shared with the root by 2. 
If  the  side in question is an internal diagonal of the original polygon   dissecting the latter 
  into two smaller  polygons, 
   choose this  side as the root of the left polygon and 
    label the edge of the $p$-valent vertex
  associated with this root by 2. 
 If  also external  side of the dissected polygon, label the
edge associated with the corner not shared with the root by 2. 
  The other possibility is that the leftmost side    is 
   a side of the original polygon. 
Then label by 2 the edge  of the $p$-valent vertex 
associated with the  corner of the side not shared with the root. 
Repeating this procedure $n$ times we  obtain a collection of $p$-valent 
vertices with numbered edges which, taken in cyclic order, 
give rise to a $p$-hyperedge diagram.
  The procedure is
illustrated in fig.  \ref{fig:dissectionplanar}.

\section{ Proof of the  Br\'ezin-Hikami contour integral  representation \re{eq:BH}}
\la{BHproof} 

The  derivation  generalises the one  given by  
Morozov-Shakirov~\cite{A.Morozov:2010ab}
for the GUE. 
For   fixed $\X$, the $\Y$-integration defines  the
Generalised Kontsevich Model partition function  
\begin{equation}
\label{eq:GKM}
  Z_{p-\mathrm{GKM}}  ( \X ) 
  \;:=\; 
  \int d \Y \, e^{  \operatorname{tr}(- \Y  \X   + {1\over p}  \Y ^p) },
\end{equation}
treated as a formal series in negative powers of $ \X $.  In
eigenvalue variables $ \X = U\,\mathrm{diag}(x_ 1,\ldots,x_
N)\,U^\dagger$, unitary invariance reduces this to a determinantal
expression:
\begin{equation}
\label{eq:GKM-det}
  Z_{p-\mathrm{GKM}}  ( \X ) 
  \;=\; 
  \kappa_N\,\frac{\det_{i,j=1,\ldots,N}\bigl[\psi_p^{(j-1)}(x_ i)\bigr]}{\Delta(x)},
\end{equation}
where $\Delta (x) = \prod_{i<j}(x_ j - x_ i)$ is the Vandermonde,
$\kappa_N$ is an $N$-dependent normalisation constant, and 
\begin{equation}
\label{eq:psi}
  \psi_p(x) \;:=\; \int_{C_p}\dd y\,\exp\!\bigl[-xy + \tfrac{1}{p}y^p\bigr]
\end{equation}
is the \emph{higher-Airy function} with $C_p$ a steepest-descent
contour ensuring convergence.  The factor $\det_{i,j}[\psi_p^{(j-1)}(x_ i)]$
is the standard $GL(N)$-invariant determinantal kernel
of the GKM   \cite{Kharchev:1991cu}.

Now we introduce an auxiliary external source $\mathrm{B} =
\mathrm{diag}(b_1,\ldots,b_N)$ on $Y$ to regularise the calculation
and take $B\to 0$ at the end.  The model with external source becomes
\begin{equation}
\label{eq:Z-extended}
  Z_p(N; s; B) \;:=\; \int dX\,dY\,\operatorname{tr}\bigl(e^{s \X
  }\bigr)\, e^{-\operatorname{tr}( \Y \X )+\operatorname{tr}( \Y ^p)/p
  + \operatorname{tr}( \mathrm{B} \X )}.
\end{equation}
Writing $ \X $ in the same eigenvalue basis as $B$  and using HCIZ 
formula we get, up to a numerical factor, 
\begin{equation}
\label{eq:Z-eigenvalue}
  Z_p(N; s; B) \;=\; \frac{1}{\Vand(b)}\,\int\prod_{i=1}^N\dd x_ i\,
  \biggl(\sum_{k=1}^N e^{s x_ k}\biggr)\, \det_{i,j}\bigl[e^{b_i x_
  j}\bigr]\,\det_{i,j}\bigl[\psi_p^{(j-1)}(x_ i)\bigr].
\end{equation}

Now write the product of determinants as a single determinant applying
Cauchy-Binet identity
\begin{equation}
\label{eq:CB}
  \int\prod_{i=1}^N\dd x_ i\,\det\bigl[f_j(x_ i)\bigr]\,\det\bigl[g_j(x_ i)\bigr]
  \;=\; 
  N!\,\det_{i,j}\!\Bigl[\int\dd y\,f_i(y)\,g_j(y)\Bigr]
\end{equation}
with $f_i(y) = e^{b_i y}$ and $g_j(y) = \psi_p^{(j-1)}(y)$.  The
insertion $\sum_k e^{sx_ k}$ in row $k$ effectively shifts the
external source value $b_k \to b_k + s$ in the $k$-th row of the
$f$-determinant.  This gives:
\begin{equation}
\label{eq:CB-result}
  \frac{Z_p(N; s; \mathrm{B} )}{Z_p(N; \mathrm{B} )}
  \;=\; 
  \sum_{k=1}^{N}
  \frac{\det_{i,j}\bigl[M_p(b_i + s\,\delta_{i,k}, j)\bigr]}
       {\det_{i,j}\bigl[M_p(b_i, j)\bigr]},
\end{equation}
where the matrix elements are
\begin{equation}
\label{eq:M-def}
  M_p(b, j) \;:=\; \int\dd y\,e^{by}\,\psi_p^{(j-1)}(y).
\end{equation}
\label{lem:M}
For each $j \geq 1$ and $b$ such that the integrals converge the
integral gives, up to a universal constant,
\begin{equation}
\label{eq:M-formula}
  M_p(b, j) \;=\; b^{\,j-1}\,e^{b^p/p}\, .
\end{equation}
 (Integrating by parts $j-1$ times,  this reduces  to $\int dy \; e^{by}\; \psi_p(y) \sim e^{b^p/p}$.)

The determinant in the denominator of \re{eq:CB-result} is
 \begin{align} 
\det_{i,j}\bigl[M_p(b_i, j)\bigr]
&= \prod_{i=1}^N e^{b_i^p/p}\,\det_{i,j}\bigl[b_i^{j-1}\bigr] 
 = \prod_{i=1}^N e^{b_i^p/p}\,\Vand(b).
\label{eq:det-collapse}
\end{align}
The shifted determinant in the numerator of~\eqref{eq:CB-result} has
$b_k$ replaced by $b_k + s$ in row $k$.  By~\eqref{eq:det-collapse}:
\begin{equation}
\label{eq:numerator}
  \det_{i,j}\bigl[M_p(b_i + s\,\delta_{i,k}, j)\bigr]
  \;=\; \prod_{i\neq k}e^{b_i^p/p}\;e^{(b_k+s)^p/p}\;
  \Vand(b_1,\ldots,b_k+s,\ldots,b_N).
\end{equation}
Combining~\eqref{eq:numerator} and~\re{eq:det-collapse}
into~\eqref{eq:CB-result} we get
\begin{equation}
\label{eq:eB-result}
  e_p(s\mid N; B) 
  \;=\;
  \sum_{k=1}^{N}
  \exp\!\Bigl[\tfrac{1}{p}\bigl((b_k+s)^p - b_k^p\bigr)\Bigr]\,
  \prod_{i\neq k}\,\frac{b_i - b_k - s}{b_i - b_k}.
\end{equation}
Now we use the identity 
\begin{equation}
\label{eq:contour-id}
  \sum_{k=1}^{N} f(b_k)\,\prod_{i\neq k}\,\frac{b_i - b_k - s}{b_i - b_k}
  \;=\;
  \frac{1}{s}\oint_{\Gamma_B}\frac{\dd u}{2\pi i}\,f(u)\,
  \prod_{i=1}^{N}\,\frac{b_i - u - s}{b_i - u},
\end{equation}
where $\Gamma_B$ is a contour encircling all $b_i$ counterclockwise.
Applying~\eqref{eq:contour-id} with $f(u) = e^{[(u+s)^p - u^p]/p}$
to~\eqref{eq:eB-result} we finally obtain 
\bee
\label{eq:contour-form}
  e_p(s\mid N; B) 
 & \;=\;
  \frac{1}{s}\oint_{\Gamma_B}\frac{\dd u}{2\pi i}\,
  \exp\!\Bigl[\tfrac{1}{p}\bigl((u+s)^p - u^p\bigr)\Bigr]\,
  \prod_{i=1}^{N}\,\frac{b_i - u - s}{b_i - u}
  \\
 &
\; \underset{b_i\to 0}
 \to\;   \frac{1}{s}\oint_{\Gamma_B}\frac{\dd u}{2\pi i}\,
  \exp\!\Bigl[\tfrac{1}{p}\bigl((u+s)^p - u^p\bigr)\Bigr]\,
  \prod_{i=1}^{N}\,\(\frac{  u +s}{ u}\)^N
\eee
 For $p = 2$ this gives the original Br\'ezin-Hikami /
 Morozov-Shakirov formula~\cite{Br_zin_2008,A.Morozov:2010ab}
\bee 
  e_2(s\mid N) 
  \;=\; \frac{e^{s^2/2}}{s}\oint_{u=0}\frac{\dd u}{2\pi i}\,e^{us}
  \Bigl(1+\frac{s}{u}\Bigr)^N,
\eee 

The multi-trace generalisation of this formula is
  \bee 
  e_p(s_1,\ldots,s_k\mid N) & \;=\;
  \prod_{i=1}^k\frac{1}{s_i}\oint\cdots\oint\prod_i\frac{\dd u_i}{2\pi
  i}\, \exp\!\Bigl[\tfrac{1}{p}\sum_i\bigl((u_i+s_i)^p -
  u_i^p\bigr)\Bigr]\, \\
 &\times  \det_{i,j}\frac{1}{u_i - (u_j+s_j)}\,
  \prod_i\Bigl(1+\frac{s_i}{u_i}\Bigr)^N.
\eee 

\section{ Proof of the Bernoulli identity \re{Bernid} }
\label{sec:n1-factor}

We evaluate the linear term in $n$ by considering $n$ as a continuous
variable and taking the derivative at $n=0$.  Denoting
\bee  P(n) := \prod_{j=0}^{2g-1}((p-1)n + 1 - j),
\quad
G_0(t) := G_n(t)\big|_{n=0} = e^t\Bigl(\frac{t}{e^t-1}\Bigr)^2
\eee
and taking into account that $P(0)=0$, we write
\begin{equation}
\label{eq:n1-factor}
  [n^1]\,\K^{(g)}_{ p}(n) \;=\; P'(0)\cdot[t^{2g}]G_0(t) \;=\;
  (p-1)\,(2g-2)!\,\bigl[t^{2g}\bigr]\,G_0(t).
\end{equation}
Since $G_0(t) $ does not depend on $p$, the $p$-dependence is entirely
through the $(p-1)$ prefactor.

Then we expand   the rhs 
using the Bernoulli generating function
\bee 
  \frac{1}{e^t-1} \;=\; \frac{1}{t}\,\frac{t}{e^t-1} 
  \;=\; \sum_{k\geq 0}B_k\,t^{k-1}/k! 
\eee 
and pick the coefficient of $t^{2g}$ to obtain
\bee 
\,\K^{(g)}_{ p}(n)  \;=\; -\frac{(p-1)\,B_{2g}}{2g}\; n\; + \; 
  \text{higher powers of } n.
\eee 
%
%
%
%
%
%
%
%
%
%
%

 \section{Tables of Higher-Genus Fuss-Catalan Numbers  }
\la{app:tableCng}

\noindent\textbf{$p =2$} \;($g_{\max}=\lfloor n /2\rfloor$):

\begin{center}
\renewcommand{\arraystretch}{1.3}
\begin{tabular}{c|cccc}
\hline
$n $ & $\FC_2 ^{(0)}(n) $ & $ \FC_2 ^{(1)}(n) $ & $\FC_2 ^{(2)}(n) $
& $ \FC_2 ^{(3)}(n) $ \\
\hline
1 & 1  & 0   & 0  & 0 \\
2 & 2  & 1   & 0  & 0 \\
3 & 5  & 10  & 0  & 0 \\
4 & 14 & 70  & 21 & 0 \\
\hline
\end{tabular}
\end{center}

\bigskip
\noindent\textbf{$p =3$} \;($g_{\max}=n $):

\begin{center}
\renewcommand{\arraystretch}{1.3}
\begin{tabular}{c|ccccc}
\hline
 $n $ & $\FC _3^{(0)}(n) $ & $\FC _3^{(1)}(n) $ & $ \FC _3^{(2)}(n) $
 & $ \FC_3 ^{(3)}(n) $ & $ \FC _3^{(4)}(n) $ \\
\hline
1 & 1  & 1    & 0     & 0       & 0     \\
2 & 3  & 25   & 12    & 0       & 0     \\
3 & 12 & 336  & 1428  & 464     & 0     \\
4 & 55 & 3630 & 51975 & 152020  & 24310 \\
\hline
\end{tabular}
\end{center}

\bigskip
\noindent\textbf{$p =4$} \;($g_{\max}=\lfloor 3n /2\rfloor$):

\begin{center}
\renewcommand{\arraystretch}{1.3}
\begin{tabular}{c|ccccc}
\hline
$n $ & $ \FC_4 ^{(0)}(n) $ & $ \FC _4^{(1)}(n) $ & $ \FC_4 ^{(2)}(n) $
& $ \FC_4 ^{(3)}(n) $ & $g_{\max}$ \\
\hline
1 & 1   & 5     & 0       & 0          & 1 \\
2 & 4   & 154   & 826     & 276        & 3 \\
3 & 22  & 2805  & 82236   & 567545     & 4 \\
4 & 140 & 41860 & 3573570 & 102782680  & 6 \\
\hline
\end{tabular}
\end{center}

 \section{A matrix model with the condition  ``restricted crossing'' imposed}
 \label{appE}
 Here we formulate a matrix  integral generating  the 
 higher genus Fuss-Catalan numbers with   the $rc$ restriction imposed.
 For that we need  $p$ pairs of hermitian matrices $ \{ X_a,Y_a\}, \ \ \ a=1,..., p$
 and replace the partition function \re{defpf}
 with
   \bee
   \la{defpfRC}
   \CZ_{N,p} = \int \prod_{a=1}^p d\X_a d\Y_a\ e^{-  S(\X,\Y)},
   \quad
     S(\X,\Y)= \sum_{a=1}^p \mathrm{tr} \X_a \Y_a - {1\over p} \mathrm{tr} [\Y_1\Y_2...\Y_p]\; .\ \ 
   \eee
 Then we have, instead of \re{Cng},   
\bee
  \sum_{g\ge 0}  \FC^{(g)}_p (n) \Big|_{ rc}\ N^{-2g} &= 
  {1\over N  ^{1+(p-1)n }}
  \< \tr  [(\X_p ...\X_2 \X_1)^n] \>_{N,p}
  \la{CngRC}
 \eee
 In the $p$-matrix integral, the order of the   
  legs of the  $p$-vertex  is frozen,
  hence  no crossing of the legs of the same vertex is allowed. 
    This model is more difficult to solve. The  $U(N)$ symmetry of the $p$-matrix integral 
  is not sufficient to reduce it as an eigenvalue integral and relative $U(N)$ rotations of the matrix variables  play   important role.
\bigskip

 
\providecommand{\href}[2]{#2}\begingroup\raggedright\endgroup

 \end{document}